\documentclass[11pt]{article}
\usepackage{amsmath,amsbsy,amsfonts,amssymb,graphics,epsfig,float,mathrsfs,amsthm,braket}
\usepackage{ifsym}
\usepackage{psfrag}
\usepackage{color}
\usepackage[normalem]{ulem}

\usepackage{graphicx}
\usepackage{amsmath}
\usepackage{amsfonts}
\usepackage{amssymb}
\usepackage{amssymb}
\usepackage{psfrag}
\usepackage{color}
\usepackage[normalem]{ulem}

\usepackage[usenames,dvipsnames]{xcolor}
\usepackage{amssymb}
\usepackage{pgfplots}
\usepackage{tikz}
\usetikzlibrary{arrows}
\usetikzlibrary{shapes.geometric}

\newcommand{\beq}{\begin{equation}}
\newcommand{\eeq}{\end{equation}}

\newcommand{\upd}{\mathrm{d}}

\newcommand{\ve}[1]{\mbox{\boldmath $#1$}}
\newcommand{\rhol}{\rho_\ell}
\newcommand{\tlift}{t_{\mathrm{LO}}}
\newcommand{\Pmin}{P_{\mathrm{min}}}
\newcommand{\ahang}{\dot{v}_{\mathrm{hang}}}
\newcommand{\vhang}{v_{\mathrm{hang}}}
\newcommand{\sclift}{s_C^{\mathrm{LO}}}

\renewcommand{\sc}{s_C}

\newcommand{\s}[1]{{\textsf{\textbf{#1}}}}

\begin{document}
\title{\s{The surprising dynamics of a chain on a pulley: Lift-off and snapping}}
\author{
\textsf{Pierre-Thomas Brun$^{1}$, Basile Audoly$^{2}$, Alain Goriely$^{3}$ and Dominic Vella$^{3}$}\\
{\it $^{1}$Department of Mathematics, Massachusetts Institute of Technology,}\\ {\it Cambridge, Massachusetts 02139, USA}\\
{\it $^2$Laboratoire de M\'ecanique des Solides, CNRS \& \'Ecole Polytechnique,}\\ {\it UMR 7649, 91128 Palaiseau, France}\\
{\it $^{3}$Mathematical Institute, University of Oxford, Oxford, OX2 6GG, UK}
}
\date{\today}
\maketitle

\hrule\vskip 6pt
\begin{abstract}
The motion of weights attached to a chain or string moving on a frictionless pulley is a classic problem of introductory physics used to understand the relationship between force and acceleration. Here, we consider the dynamics of the chain when one of the weights is removed and, thus, one end is pulled with constant acceleration. This simple change has dramatic consequences for the ensuing motion:  at a finite time, the chain `lifts off' from the pulley and the free end subsequently accelerates \emph{faster} than the end that is pulled. Eventually, the chain undergoes a dramatic reversal of curvature reminiscent of the crack, or snap, of a whip. We combine experiments, numerical simulations, and theoretical arguments to explain key aspects of this dynamical problem.
\end{abstract}
\vskip 6pt
\hrule

\maketitle

%
%

\section{Introduction}

When publishing the design of his machine in 1784, little could Atwood have known that,  more than two centuries later,  students would be asked to predict the outcome of his experiment. The problem, traditionally offered  as an illustration of the principles of Newtonian mechanics, consists in deriving the acceleration $\ve a$ of two masses $M>m$ subject to the action of gravity $\ve g$ while attached to a massless and inextensible chain passing over a frictionless pulley. The well-known result is that
\beq
\ve a =  \frac{M-m}{M+m} \ve g\,
\label{eqn:AtwoodAccel}
\eeq
for the heavier of the two, with the second having the opposite acceleration. Little attention has been paid to the seemingly trivial case in which both $m=0$ and the chain has a finite linear density. If the mass of the chain nevertheless remains small compared to $M$, then equation \eqref{eqn:AtwoodAccel} immediately  gives that the remaining mass falls with acceleration $a=g$. However, a simple experiment (see, for example, figure \ref{fig:setup}b) reveals that this  apparent simplification actually has a dramatic effect on the resulting motion: the chain `lifts off' from the pulley in a complex motion.  This lift-off has some surprising features that we explore in detail in this paper. In particular, we show that the free end accelerates \emph{faster} than the end that is being pulled by the mass: in this sense, the free end `beats' the free fall of the mass. We also show that the chain eventually `snaps' in a manner that is reminiscent of the crack of a whip\cite{goriely2002shape,mcmillen2003whip}.  

Thin  filamentary structures are as important in applications as they are ubiquitous in nature and industry; examples range from macromolecules such as DNA~\cite{Neukirch:2004cu} to the kilometric transoceanic cables laid on the ocean bed from telecommunication vessels~\cite{carter2009submarine}. On a human scale, textiles, hair and ropes are other examples of thin elongated structures in which one dimension greatly exceeds the two others (so that the objects may be modeled as one-dimensional). The thinness of such rod-like structures also makes them flexible so that they are frequently subject to large deformations in the three dimensional environment in which they evolve. This flexibility in shape in turn leads to rich bifurcation landscapes~\cite{brun2014introduction}, striking pattern formation~\cite{habibi2007coiling}, and intricate dynamical behavior.

Despite the range of length scales and materials encountered, many filamentary structures are well modeled by the Kirchhoff equations for elastic rods~\cite{audoly2010elasticity}. A particularly interesting limit of these equations is the case of inextensible strings in which inextensible rods have  negligible resistance to bending and twisting:  their  behaviour is dictated to a large extent by the geometric constraint of inextensibility. In fact, this constraint by itself is enough to give rise to complicated dynamics: while the simplest case of a straight string accelerating along its length may be understood by a simple application of Newton's second law, any closed shape is a solution of the governing equations with any constant tangential velocity \cite{Healey1990}. It is therefore the combination of acceleration and `turning a corner' that gives rise to the most interesting dynamics. In particular, moving inextensible strings  form surprisingly complicated shapes including arches~\cite{hanna2012slack} and the mesmerising `chain fountain'~\cite{biggins2014understanding,biggins2014epl}. 

In this paper we consider the planar motion of a chain moving around a pulley subject to a constant acceleration at one end and free at the other end. We model the chain as an inextensible string in partial contact with a disk (the pulley) and investigate key features of the motion. This setup is perhaps the simplest geometry in which one can study how a string or chain `turns a corner' since the curvature of this corner is simply that of the pulley. We first describe in some detail the different phenomena that are observed experimentally and then analyze  the lift-off in detail by combining theoretical arguments with  numerical simulations.  Finally, we describe the  snapping that is associated with the  reversal of curvature that ultimately occurs close to the free end.

\section{Statement of the problem and experimental observations}
\label{phenomenon}

\begin{figure}
\centering
\includegraphics[width=\textwidth]{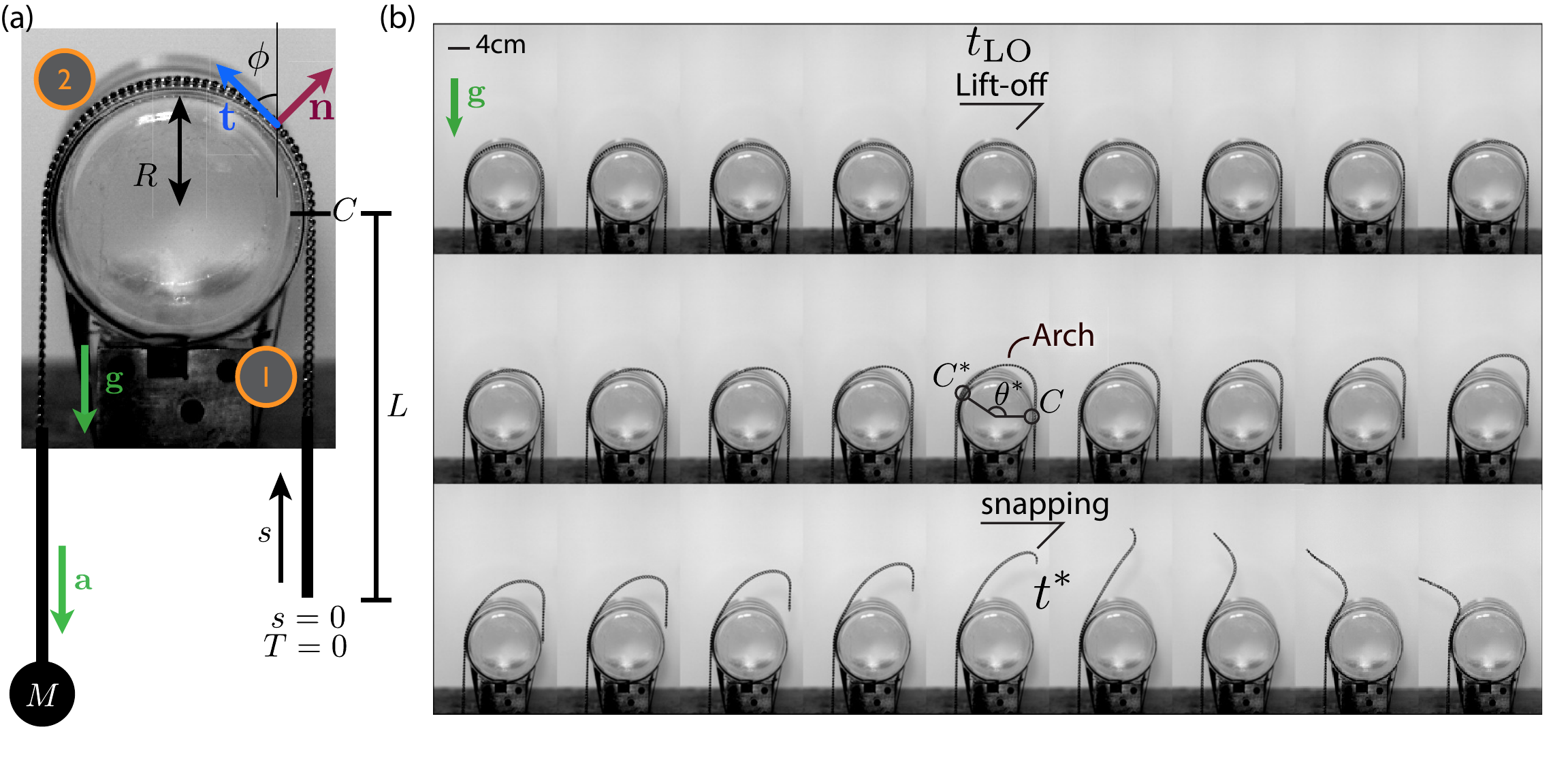}
\caption{ Experiments using ball chains. (a) Experimental setup: the free fall of a mass $M$ forces a
chain with initial hanging length $L=54\mathrm{~cm}$ to slide around a
glass cylinder of radius $R=4\mathrm{~cm}$.  The initial
length $L$ of the hanging part denotes the section of chain from the
free end $s=0$ to the first contact point with the cylinder $C$ before
the mass is dropped (which occurs at $t=0$).  The vectors $(\ve t, \ve
n)$ denote the tangent and the normal of the chain, respectively.  (b)~Time sequence ($\Delta t=2.5\mathrm{~ms}$) of the chain dynamics
successively lifting up from pulley (from $t=\tlift$ onwards) showing
that a ballooning arch develops before the chain eventually `snaps' at
a time $t=t^*$ (reverses its curvature dramatically).  }
\label{fig:setup}
\end{figure}

In the idealized setup of our problem, a mass $M$ is attached to the end of a chain of linear density (mass per unit length) $\rhol$ placed around a circular pulley of radius $R$. The mass $M$ is released at time $t=0$ and subject to a constant acceleration $a$.   The chain  is initially held on the pulley with a known length  of chain, $L$, hanging free from the first point of contact $C$ (see Fig~\ref{fig:setup}a). In a typical experiment the total length of the chain is $L_\text{end}=2 L+\pi R$. 

The realization of this thought-experiment was  carried out by using conventional ball and link chains purchased from the local do-it-yourself shop and hung over horizontal Pyrex beakers (Fisher Scientific) that were used as the frictionless pulley (the surface of the beaker being  smooth, the chain slides over the surface with very little friction). The mass $M{\gg \rhol L}$ is released at time $t=0$ resulting in an acceleration $a\approx g$. Although not restricted in any other way than by a frictionless contact with the pulley, the motion of the chain was observed to be planar, $i.e.$~the motion remains in the $xy-$plane normal to the axis of the pulley and containing the chain at $t=0$. 
In the subsequent modelling, we therefore assume that the motion remains planar throughout.

Typically, the time evolution of the chain passes through three, qualitatively different, phases shown in figure  \ref{fig:setup}b: 
\begin{itemize}
\item For $0\le t<\tlift$ the chain follows the dynamics that might
naively be expected: it moves around the pulley at the speed that is
imposed by the accelerating mass.  

\item At a time $t=\tlift$ the chain starts to partially lose contact
with the pulley, lifting off so that for $t>\tlift$ the hanging part
of the chain goes faster than the rest of the chain.  Due to the
excess length of the chain, an arch forms between points $C$ and
$C^*(t)$ defining an angle $\theta^*(t)$ (see figure
\ref{fig:setup}b).

\item After the free end goes past the last contact point $C$ (first
image of the last row in figure \ref{fig:setup}b) the arch flares up,
its curvature increasing rapidly until the free end eventually snaps
at time $t=t^{*}$ (the curvature there changes sign).

\end{itemize}

To complement our experiments, numerical simulations were carried out
using the Discrete Elastic Rod method, which provides a discretization
of the equations of motion for thin elastic rods using a Lagrangian
formulation~\cite{bergou2008discrete}.  This method, and its
counterpart for thin viscous threads, have been used and validated
previously~\cite{Jawed:2014gs,Audoly:2013cb}.  In the simulations, we
use a string model, i.e.~the bending and twisting stiffnesses are both
set to zero so that the Kirchhoff equations~\cite{audoly2010elasticity} simplify to the conservation of linear momentum:
\beq
\label{eqn:KirchhoffR}
\frac{\partial (T \ve t) }{\partial s} + \rhol \ve g+  P \ve n= \rhol \frac{\partial \ve v }{\partial t},
\eeq and inextensibility
\beq \left | \frac{\partial \ve x (s,t)}{\partial s} \right |  =1,
\label{eqn:KirchhoffR2}
\eeq where $\ve x (s,t)$ is the centre-line of the chain as a function
of the arc length $s$ and time $t$, $\ve{t}(s,t) = \frac{\partial \ve
x (s,t)}{\partial s}$ and $\ve{n}(s,t)$ are the unit tangent and
normal vectors to the chain axis, respectively (see
Figure~\ref{fig:setup}a).  Further, the variables $T(s,t)$ and $\ve
v(s,t) = \frac{\partial \ve x (s,t)}{\partial t}$ denote the tension
and velocity of the chain, and $P(s,t)$ represents the (frictionless)
reaction from the pulley on the chain, such that $P=0$ whenever there
is no contact.  This frictionless contact with the pulley is
implemented numerically using a geometrical method, \emph{i.e.}\ by
alternating dynamic steps that ignore contact forces, with projection
steps in which the configuration is projected onto the manifold of
admissible configurations.

We close the differential problem (\ref{eqn:KirchhoffR})-(\ref{eqn:KirchhoffR2}) using the boundary
 conditions at the two ends of the chain. Namely, the free end of the string is stress free and  we prescribe the acceleration $a$ of the chain's end attached to the mass in our experiments, that is: 

\begin{align}
&T(0,t)=0 \\
&\dot {\ve v }(L_\text{end},t)= -a\,  \ve e_y
\label{eq:bc}
\end{align}
 where $\ve{e}_y$ is in the vertical direction and $L_\text{end}=2L+\pi R$ is the arc-length corresponding to the point where the mass is attached.  
\begin{figure}
\centering
\includegraphics[width=.97\textwidth]{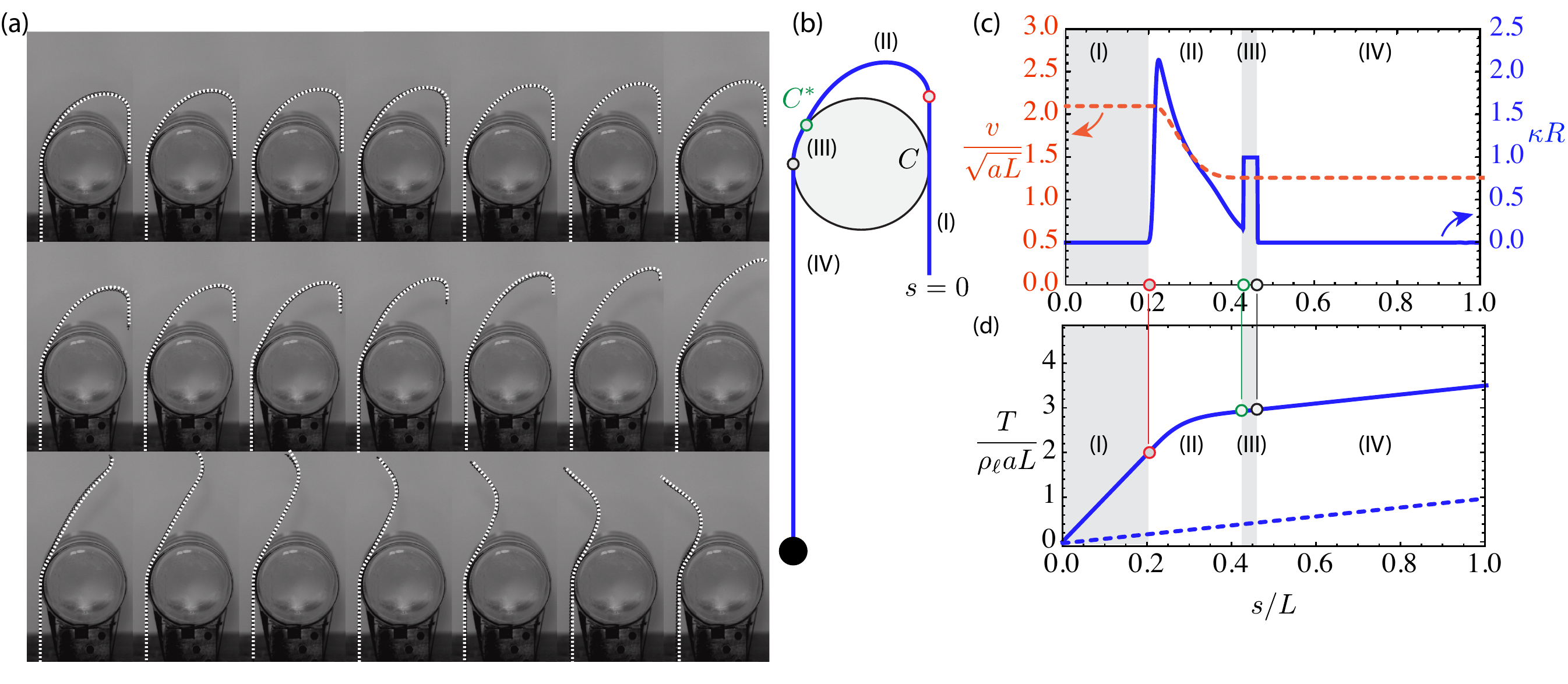}
\caption{Numerical simulations: (a) Comparison between the chain shape observed
experimentally (images in the background) and the prediction of
numerical simulations (overlaid dashed curve).  Here there are no
adjustable parameters ($L=54\mathrm{~cm}$, $R=4\mathrm{~cm}$, time
between two frames $\Delta t=0.9\mathrm{~ms}$).  (b) At each instant
before snapping, the chain may be divided into four zones.  (c) Plot
of the curvature $\kappa$ (dashed curve) and the chain speed $v$
(solid curve) as functions of the arc length.  (d) A numerical
reconstruction of the tension $T$ within the chain shows that it
varies linearly with arc length in regions (I), (III) \& (IV).  Region
(II) bridges regions (I) and (III), each of which is characterized by
a different acceleration.  For comparison, the dashed line shows the
tension as a function of arc length in the idealized solution with
uniform and constant acceleration $a$ where lift-off is prevented
(this artificial solution would give rise to a negative contact
pressure).}
\label{fig:DVR}
\end{figure}
Figure~\ref{fig:DVR} shows a direct comparison between experiments and
numerical simulations without any adjustable parameter.  The favorable
agreement between the two validates the approximations made in
modelling the chain as a string that is driven at constant
acceleration (our experiments are in fact conducted with a constant
force $M \ve g$ but the two methods are approximately equivalent since
$ \rho_\ell L \ll M$).  Of particular interest is the possibility
offered by the simulations to access the successive space derivatives
of the basic physical variables (such as curvature) with controllable
time resolution covering long periods of time.  Additionally, physical
quantities that are otherwise challenging to measure experimentally,
such as the tension within the chain, are readily available from
simulations.

At any given time before the chain snaps ($0<t<t^*$), the chain may be
divided into four regions: (I) and (IV) denote the two straight,
vertical parts, respectively, (II) denotes the part of the chain not
in contact with the pulley and (III) the contact region.  For example,
in figure \ref{fig:DVR}b-d, only a small portion of the chain (III)
remains in contact with the pulley.  In this region the chain's speed
is prescribed by the free fall of the mass, as in region (IV).  In
region (I), the chain moves significantly faster than the imposed
acceleration, while remaining perfectly straight and tangential to the pulley.  In addition, the tension in the chain is larger than if the lift-off were artificially suppressed (shown as a dashed line
in figure \ref{fig:DVR}d).  The area (II) bridges the areas (I) and
(III)-(IV) and their corresponding physical quantities, resulting in a
ballooning shape (see figure \ref{fig:DVR}).

The length of the  chain that is initially hanging freely, $L$, turns out to play an important role in the problem and, therefore, we use $L$ as the length scale of the problem. It is then natural to use the quantity $(L/a)^{1/2}$ as the time scale of the problem. If the acceleration of the freely hanging part of the chain were to remain at $a$ throughout the motion, the length of the chain hanging beneath the point C then shrinks to zero at a time $t=\sqrt{2L/a}$.

From the 6 dimensional quantities of interest, we identify three
dimensionless parameters: \beq \pi_1= \rhol L / M, \,\,\,\,
\pi_2 = R/L, \,\,\,\, \text{and} \,\,\,\, \pi_3 = g/a.
\label{eqn:DimlessPars}
\eeq In our experiments $M$ is chosen such that $\pi_1 \ll 1$; $\rhol$
therefore enters in the problem as a multiplying factor for forces
only.  The value of $\pi_2 $ may easily be varied experimentally by
varying the radius of the pulley or, more simply, by varying the
length of chain that initially hangs freely.  By contrast, varying
$\pi_3$ is more difficult.  Our experiments are always performed with
$a=g$ and hence $\pi_3 = 1$.  However, the inclusion of a body force
acting on the chain complicates some of the analysis without changing
the qualitative behaviour, as shown in figure \ref{fig:DVR}.
Therefore, in much of the analysis we shall neglect the acceleration
due to gravity, i.e.~we assume $\pi_3=0$.  Experimentally, this case
could be obtained by placing the chain, pulley and the mass on a
smooth horizontal surface: at $t=0$, the mass is then pushed off the
table into the air~\cite{hanna2014jump} and allowed to fall under
gravity so that the acceleration of the chain is $a=g$ but the
effective gravity acting directly on the chain vanishes.  In the
simulations, the value of $\pi_3$ is set to 1 or 0 depending on whether
the results are to be compared with experiments or with our analysis.


\section{Prior to lift-off\label{sec:b4lift}}

We consider a pulley of radius $R$ around which a chain with linear density $\rhol$ is hung.  We begin by including gravity as a body force, though later we will neglect it to simplify the ensuing calculations.
Initially, the length of chain between the free end (at
arc-length position $s=0$) and the chain's first point of contact with
the pulley is $L$ (i.e.~the point C in figure  \ref{fig:setup}  has arc-length coordinate $s=L$ at $t=0^-$).

At $t=0^+$, the end at arc length $s=L_\text{end}$ is pulled with
constant acceleration $a$.  Assuming that the chain remains in contact
with the pulley, $\ve v = v \ve t$, and
projecting \eqref{eqn:KirchhoffR} onto the tangential and normal
directions, and then using the Frenet formula $\partial\ve{t}/\partial
s=-\kappa\ve{n}$ with $\kappa$ the curvature, we find that
\begin{eqnarray}
\label{k1}
\frac{\partial T}{\partial s} - \rhol g\cos\phi & = &  \rhol \frac{\partial v}{\partial t} \\
\label{k2}
-{T\kappa}- \rhol g\sin\phi+ P & = & -\rhol {v^2\kappa} ,
\end{eqnarray} where $\phi$ is the angle between the local tangent to the chain and the upward pointing vertical and $\kappa=\partial\phi/\partial s$ is the curvature.  In particular, while the chain remains in contact with the pulley, $t<\tlift$, we know that
\begin{align}
\phi(s,t) =\begin{cases} 0,& 0\leq s<\sc\\
 \frac{s-\sc}{R},& \sc\leq s<\sc+\pi R\\
 \pi,& \sc+\pi R\leq s
 \end{cases}
 \label{eq:phiBeforeLO}
\end{align} where 
\beq
\sc(t)= L-at^2/2
\label{eqn:scDefn}
\eeq is the arc-length position of the first contact point between the pulley and the chain (the point labelled C in figure  \ref{fig:setup}). In the hanging part $\phi(s,t)=0$, and {so, integrating \eqref{k1} subject to $T(0,t)=0$, we find that} the tension $T$ may be written:
\beq
\label{eq:tension}
T(s,t) = \rhol (a +g ) s  ,  \,\,\, s \in [0,\sc]
\eeq
In the region where the chain is in contact with the pulley, \eqref{k1} leads to
\beq
T(s,t) =  \rhol a s +\rhol g R \sin \phi + C_1, \,\, s\in[\sc,\sc +\pi R]
\eeq where the constant of integration $C_1=  \rhol g \,\sc$ is found by requiring the tension to be continuous at $s=\sc$.

{In the freely hanging portion of the chain, $\eqref{k2}$ is identically satisfied by $P=0$. Elsewhere, \eqref{k2} can be viewed as an equation for the reaction of the pulley on the chain required for this motion to occur. From a simple rearrangement, we  find that}
\beq
P = -\rhol \frac{v^2}{R} + 2  \rhol g \sin\phi +  \frac{1}{R}\rhol ( a s + g\sc), \quad s\in[\sc,\sc +\pi R].
\eeq Note that $P$ is a function of arc length $s$ and time $t$
through the angle $\phi$ and arc length of the contact point,
$\sc(t)$, in addition to the explicit dependence on $s$.  The minimum
value of $P$ at any particular time $t<\tlift$ is attained at
$s=\sc$: \beq \Pmin(t)= \rhol\bigl[ \sc ( a+ g)- v^2
\bigr]/R=\rhol\bigl[ ( a+ g)L- \tfrac{1}{2}at^2(3a+g) \bigr]/R.
\label{eqn:Pmin}
\eeq

The key to understanding why lift-off occurs is held by the expression
for the smallest reaction force $\Pmin(t)$ \eqref{eqn:Pmin}.  At
sufficiently early times $\Pmin>0$, and the contact pressure is
everywhere positive (it prevents the chain from penetrating the
pulley).  However, $\Pmin$ vanishes at a time $t=\tlift$ 
given by
\beq
\label{eq:tlo}
\tlift=\sqrt{\frac{2L(a+g)}{a(3a+g)}}
=\sqrt{\frac{2\,L}{a}}\sqrt{\frac{1+\pi_3}{3+\pi_3}},
\eeq
and then becomes negative.  For time $t>\tlift$, the
assumption~(\ref{eq:phiBeforeLO}) that the string is in contact with
the pulley over half a circle breaks down, as it predicts a negative
pressure (an adhesive force would be required to maintain contact).  The
negative values of $P$ are first attained in the neighbourhood of $C$,
which suggests that lift-off takes place there.  At time $t=\tlift$,
the length of the hanging chain is \beq
\sclift=\sc(\tlift)=\frac{2a}{3a+g}L=\frac{2}{3+\pi_3}L.
\label{eqn:HangingLengthLO}
\eeq 

In \eqref{eq:tlo}, $\tlift$ is less than the time $\sqrt{2L/a}$ when
the hanging part (I) shrinks to a point the mathematical
solution~(\ref{eq:phiBeforeLO}) ignoring the lift-off.  This confirms
that the hanging part (I) has a finite length when the lift-off takes
place. This is also consistent with $\sclift = \sc(\tlift) >0$.

Apart from factors obviously required by dimensional analysis, both
\eqref{eq:tlo} and \eqref{eqn:HangingLengthLO} appear to depend only
on the ratio $\pi_3=g/a$ of the acceleration due to gravity to the
imposed acceleration. They do not depend on the dimensionless group
$\pi_2=R/L$ involving the size of the pulley.  Furthermore, we note
that in the limit of zero gravity, $\pi_3=0$, one third of the
original hanging chain has passed around the pulley with two-thirds
still hanging; when pulling the chain using a heavy mass in free fall,
i.e.~imposing $\pi_3=1$, then one-half of the original hanging chain
has passed around the pulley with one-half still hanging.  Here,
incorporating gravity makes a quantitative, but not qualitative,
difference; either way, an appreciable fraction of the initially
hanging chain remains below the first contact point C.

From the above argument, we also gain some intuition into \emph{why} the chain lifts off: the tension at the contact point $\sc(t)$ is decreasing with time (because the length of hanging chain is decreasing) while the tension required to turn the chain  around the pulley increases (because the chain is moving faster and faster). Lift-off occurs when the excess tension in the chain due to the hanging length is `used up'. 

\section{Lift-off}

\begin{figure}
\centering
\includegraphics[width=\textwidth]{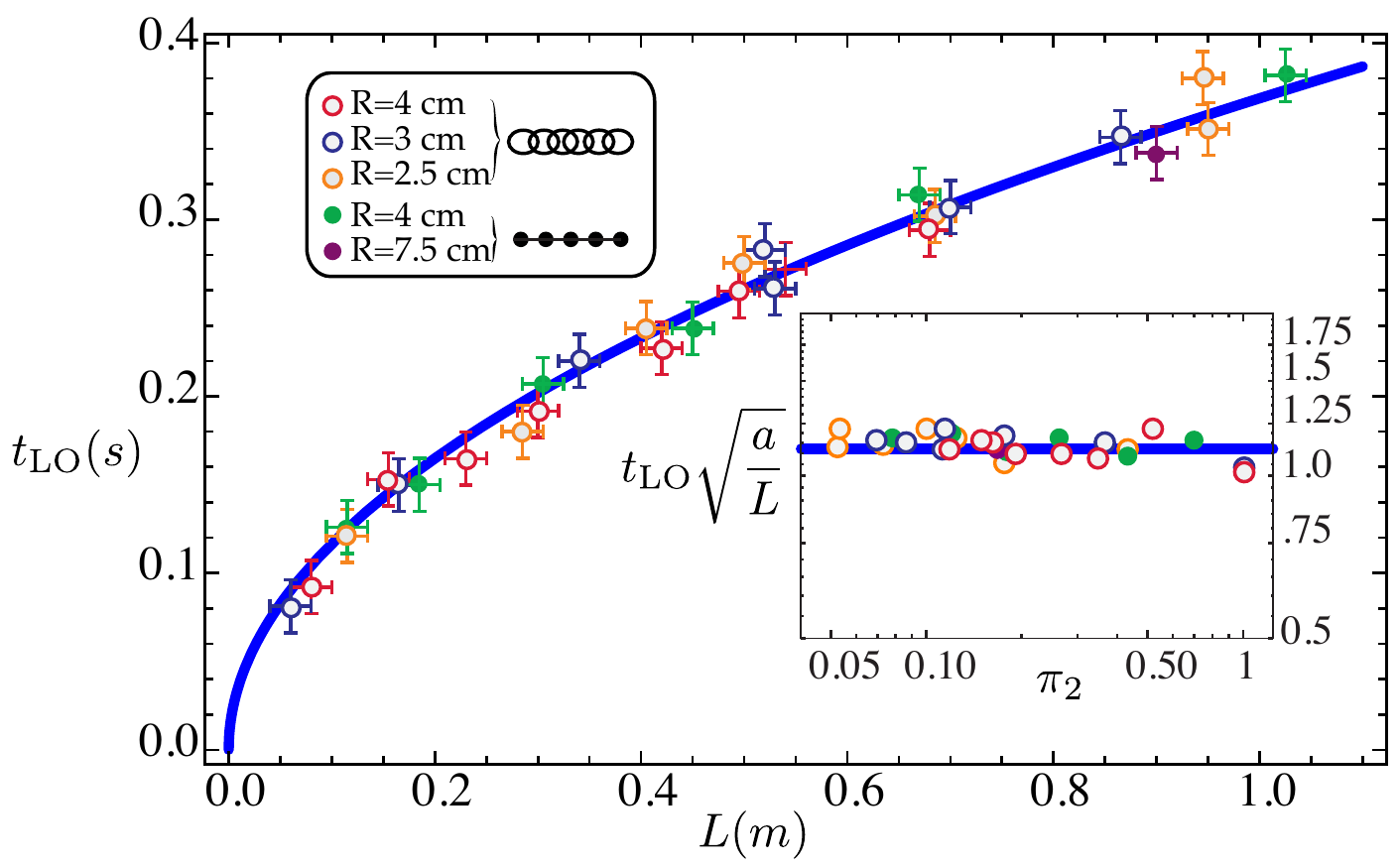}
\caption{Experimental values of the time at which lift off occurs, $\tlift$, plotted as a function of the chain length $L$.  This data, gathered with pulleys of radius ranging from $R=2.5\mathrm{~cm}$ to $R=7.5\mathrm{~cm}$ and two types of chain, collapses on a single master curve. The prediction derived in \S \ref{sec:b4lift} (see eq.~(\ref{eq:tlo})) is verified within $10\%$ (the solid curve corresponds to $\tlift\simeq1.1\sqrt{L/g}$) . {In these experiments, $a=g$, i.e.~$\pi_3=1$.}}
\label{fig:LOdata}
\end{figure}

The prediction for the lift-off time, \eqref{eq:tlo}, presents a
natural test of the model assumptions, in particular with regard to
our neglect of the bending stiffness of the chain and the assumption
of a frictionless contact between chain and pulley.  Experiments were
performed with $a=g$ (i.e.~$\pi_3=1$) and a number of different pulley
sizes and initial hanging lengths (i.e.~varying $\pi_2$).  As shown in
figure \ref{fig:LOdata}, we find reasonable agreement between the
theoretical prediction for $\tlift$ and that measured experimentally.
In particular, experiments confirm the result that the size ratio
$\pi_2=R/L$ does not affect the onset of lift-off.  We also found that
the value of $\tlift$ is insensitive to the type of chain used (ball
chains and link chains give the same results).

However, we note that the onset of lift-off is difficult to assess
accurately in experiments owing to its slow initial dynamics (to be
discussed in detail in \S~\ref{after}); it is therefore natural that
experimental values of $\tlift$ are slightly larger that theoretically
predicted (though they remain within 10\% of the theoretical value).
Furthermore, the favorable agreement between experiments and numerical
solution of the full model, which neglects any source of dissipation,
shown in figure \ref{fig:DVR}a suggests friction cannot be the cause
of the discrepancy.


\section{Behaviour just after lift-off}
\label{after}
We now turn our attention to the early phase of the motion, just after lift-off. In this phase, our experiments and numerical simulations show that the chain `balloons' off the pulley (see figure  \ref{fig:setup}b). For simplicity, we shall henceforth neglect the role of gravity, i.e.~we set $\pi_3=0$. This analysis is conducted hand in hand with numerical simulations for $\pi_3=0$, but  these results should not be compared with experiments (for which $\pi_3=1$). In Appendix \ref{sec:grav} we consider how the results of this section change when $\pi_3\neq0$; this analysis shows that the presence of gravity makes quantitative, rather than qualitative, changes to the motion around lift-off.

\subsection{Velocity after lift-off\label{sec:aftervel}}

\begin{figure}
    \begin{center}
	\includegraphics[width=\textwidth]{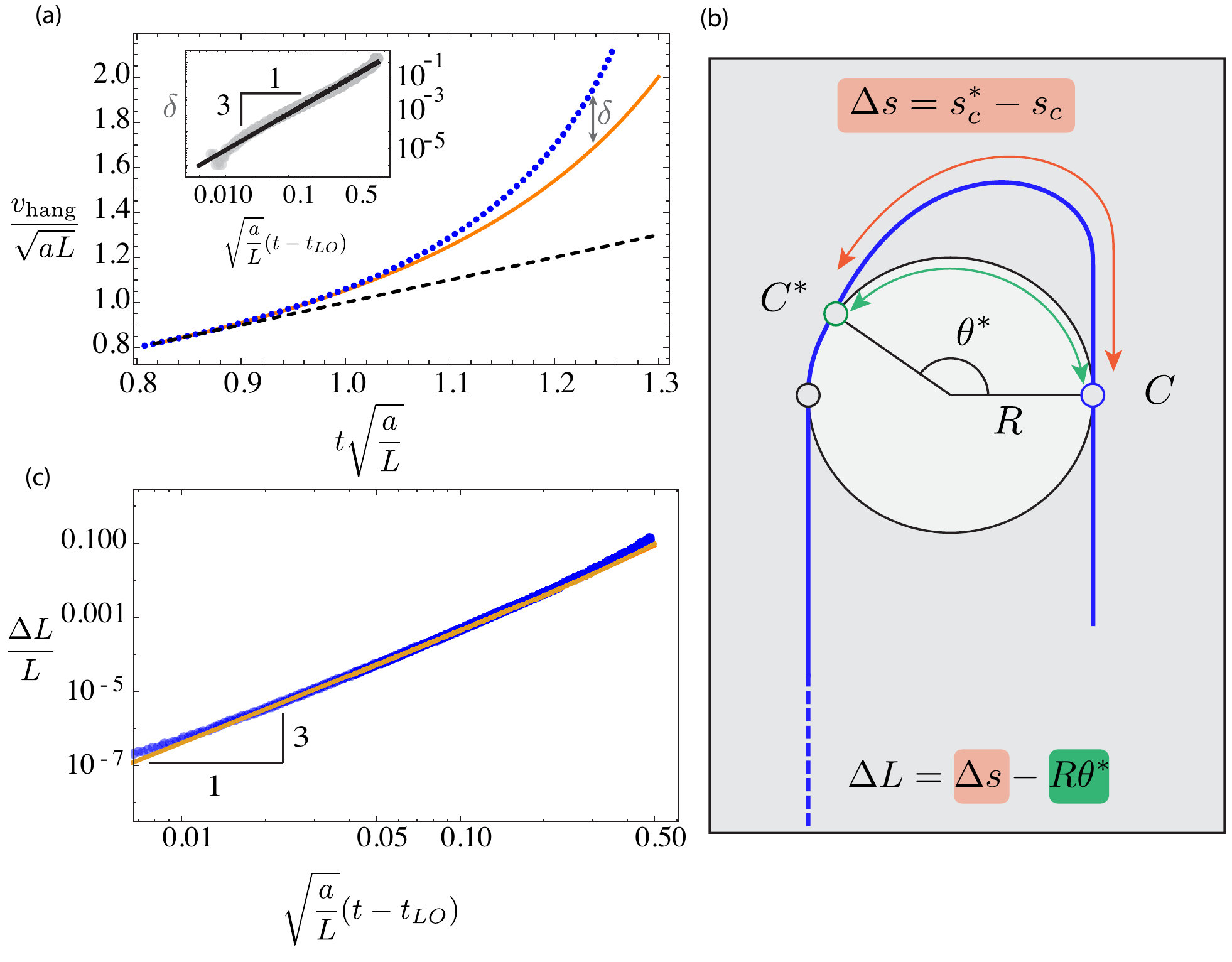}
	\caption{Behavior just after lift-off for $\pi_3=0$ (imposed
	acceleration at endpoint is $a=0.5$, $g=0$, $R=1$ and
	$L=4\pi$; the simulation timestep is $\Delta
	t=15\times10^{-4}$ and the mesh size is $\Delta s =
	6.3\times10^{-3}$).  (a) The speed of the hanging part of
	the chain for $g=0$ from numerical simulations (dotted blue
	curve) compared to the imposed velocity at the pulling end,
	$v=at$, (dashed line) and the prediction of
	eq.~\ref{eqn:vhanggeq0} (solid orange curve).  Inset: Error
	$\delta$ between the estimate \eqref{eqn:vhanggeq0} for
	$\vhang$ and the numerical data{, shows that the expansion for
	$t-\tlift\ll\tlift$ in \eqref{eqn:vhangtaylor} is correct at
	$O(t-\tlift)^2$}.  (b) $\Delta s$ denotes the length of chain
	that has lost contact with the pulley for $t>\tlift$.  The
	corresponding angle is $\theta^*$ representing a region on the
	pulley of length $R\theta^*$.  The excess length $\Delta L$ is
	defined as the difference between these two quantities $\Delta
	L = \Delta - R\theta^*$ (c) The value of {the relative excess
	length} $\Delta L/L$ (solid) as predicted by equation
	\eqref{eqn:DeltaL} compares favourably with numerical data
	(dots); in particular the scaling prediction $\Delta L/L\sim
	(t-\tlift)^3$ is borne out by the data.}
	\label{hsm}
    \end{center}
\end{figure}

The numerical simulations of the complete theoretical model (eqs~\ref{eqn:KirchhoffR}-\ref{eq:bc}) and experiments both suggest that the
portion of the chain {that is freely hanging at the instant of first
lift-off} (i.e.~$0\leq s<\sc(\tlift)$) remains straight (within numerical accuracy)
in the immediate aftermath of lift-off.  We therefore attempt to
repeat the calculation of \S\ref{sec:b4lift} but this time we treat
the acceleration in this straight portion, $\ahang(t)$, as an unknown.
Motivated by numerical observations, we assume that the length of the
straight portion of the chain is precisely the hanging length of the
chain at $\tlift$, namely $\sc(\tlift)$, which we denote by $\sclift$ 
as in \eqref{eqn:HangingLengthLO}.

Setting $g=0$ in \eqref{eqn:KirchhoffR}, and focusing on the straight, hanging part (zero curvature), and integrating we find that:
\beq
T=\rhol \ahang s, \quad 0\leq s\leq \sclift
\label{eqn:tensaccel}
\eeq where the length of the straight portion $\sclift=L-\tfrac{1}{2}a\tlift^2$. 
To close the problem, we note that at the end of this straight portion, the radius of curvature must change to some finite value (by construction), and further that there is no reaction force (since at this point the chain is no longer  in contact with the pulley). Substituting these results into \eqref{k2}, we then have that $T(\sclift,t)=  \rhol \vhang^2$ and hence, after combining with \eqref{eqn:tensaccel}, that
\beq
\ahang(t)=   \frac{\vhang^2(t)}{\sclift}.
\label{eqn:HSAnew}
\eeq 

\noindent Integrating \eqref{eqn:HSAnew} subject to the initial condition $\vhang(\tlift)=a\tlift$, we find that
\beq
\vhang(t)=\frac{2}{3}\,\frac{L}{2\tlift-t} = 
a\,t_{\mathrm{LO}}\,\left(2-\frac{t}{t_{\mathrm{LO}}}\right)^{-1}
\textrm{.}
\label{eqn:vhanggeq0}
\eeq This simple approximation predicts that the velocity of the
hanging portion of the chain diverges as $t\to2\tlift$.  {However, the
argument above only holds in the immediate aftermath of lift-off,
$t-\tlift\ll\tlift$.  In the simulations, the reversal of curvature
actually occurs at a time $t^* \approx
1.365\sqrt{L/a}\approx1.67\tlift$ (see figure \ref{snap}b-c), which is
earlier than that predicted by~\eqref{eqn:vhanggeq0}.  

The Taylor expansion of \eqref{eqn:vhanggeq0} for $t-\tlift\ll\tlift$
reads \beq
\vhang(t)=a\,t_{\mathrm{LO}}\,\left[1+\frac{t-\tlift}{\tlift}+\left(\frac{t-\tlift}{\tlift}\right)^2+O\left(\frac{t-\tlift}{\tlift}\right)^3\right].
\label{eqn:vhangtaylor}
\eeq The first two terms in in the square bracket combine to give
$at$, which is the imposed pulling velocity: the correction to the
imposed constant acceleration is given by the following term and
occurs at order $(t-\tlift)^2$.  In figure \ref{hsm}a, we observe that
our numerical results for the velocity are in good agreement with the
prediction of eq.~\eqref{eqn:vhanggeq0} for times close to $\tlift$
but then deviate with an error $\delta$ that is cubic in $t-\tlift$.
This cubic error confirms that the first correction to the constant
acceleration motion, namely the term of order $(t-\tlift)^2$ in
\eqref{eqn:vhangtaylor}, is correct.  }

By differentiating \eqref{eqn:vhanggeq0} with respect to time, we obtain an expression for the excess acceleration of the free part of the chain
\beq
\Delta a=\ahang-a=a\left[\frac{\tlift^{2}}{(2\tlift-t)^{2}}-1\right],
\label{eqn:deltaa}
\eeq which is  positive for all $t>\tlift$. Thus, even though the free end is not subject to external forces, it accelerates \emph{faster} than the end that is actually subject to the imposed (constant) acceleration.

A  quantity of interest is the arc-length of the region over which the chain has lost contact with the pulley for $t>\tlift${, which we denote by $\Delta s$ (see figure \ref{hsm}b}). The above calculation does not give us enough information to determine this quantity. However, the result for $\vhang$, \eqref{eqn:vhanggeq0},  can be used to determine the excess length of chain that has lost contact with the pulley, i.e.~the difference between the length of chain that has lifted off the pulley, $\Delta s(t)$, and  the arc length of the pulley from which it has lifted off, $R\theta^*(t)$, with $\theta^*(t)$  the angle subtended between the two contact points C and C$^*$ (see figure \ref{hsm}b). Denoting this quantity by $\Delta L(t)=\Delta s(t)-R\theta^*(t)$ we find that
\beq
\Delta L(t)=\int_{\tlift}^t\left[\vhang(t')-at'\right]~\upd t'= \frac{2L}{3}  \log \left(\frac{\tlift}{2\tlift-t}   \right)-\frac{a}{2}(t^2-\tlift^2).\label{eqn:DeltaL}
\eeq 
We emphasize that this calculation does not allow us to determine $\Delta s$ and $R\theta^*$ separately, but only their difference.
The Taylor series of $\Delta L(t)$ for $t-\tlift\ll\tlift$ reveals that
\beq
\label{approx}
\Delta L(t)\approx a{(t-\tlift)^3\over 3\tlift}, \eeq \emph{i.e.}~the
excess length grows only relatively slowly after the start of
lift-off.  Note that the appearance of a leading-order behaviour at
order $(t-\tlift)^3$ here is consistent with our earlier finding that the
correction to the speed of motion is at order $(t-\tlift)^2$ in
\eqref{eqn:vhangtaylor}, since we have integrated this quantity in
time.

Comparison of the predictions of this analysis, \eqref{eqn:vhanggeq0}
and \eqref{approx}, with the numerical solution of the fully nonlinear
problem (see figure \ref{hsm}) shows that these expressions are
asymptotically correct immediately after lift-off,
$t-\tlift\ll\tlift$, and validates the underlying assumption (that the
hanging portion of the chain simply accelerates vertically upwards).

\subsection{Characteristics}

\begin{figure}
\begin{center}
\includegraphics[width=\textwidth]{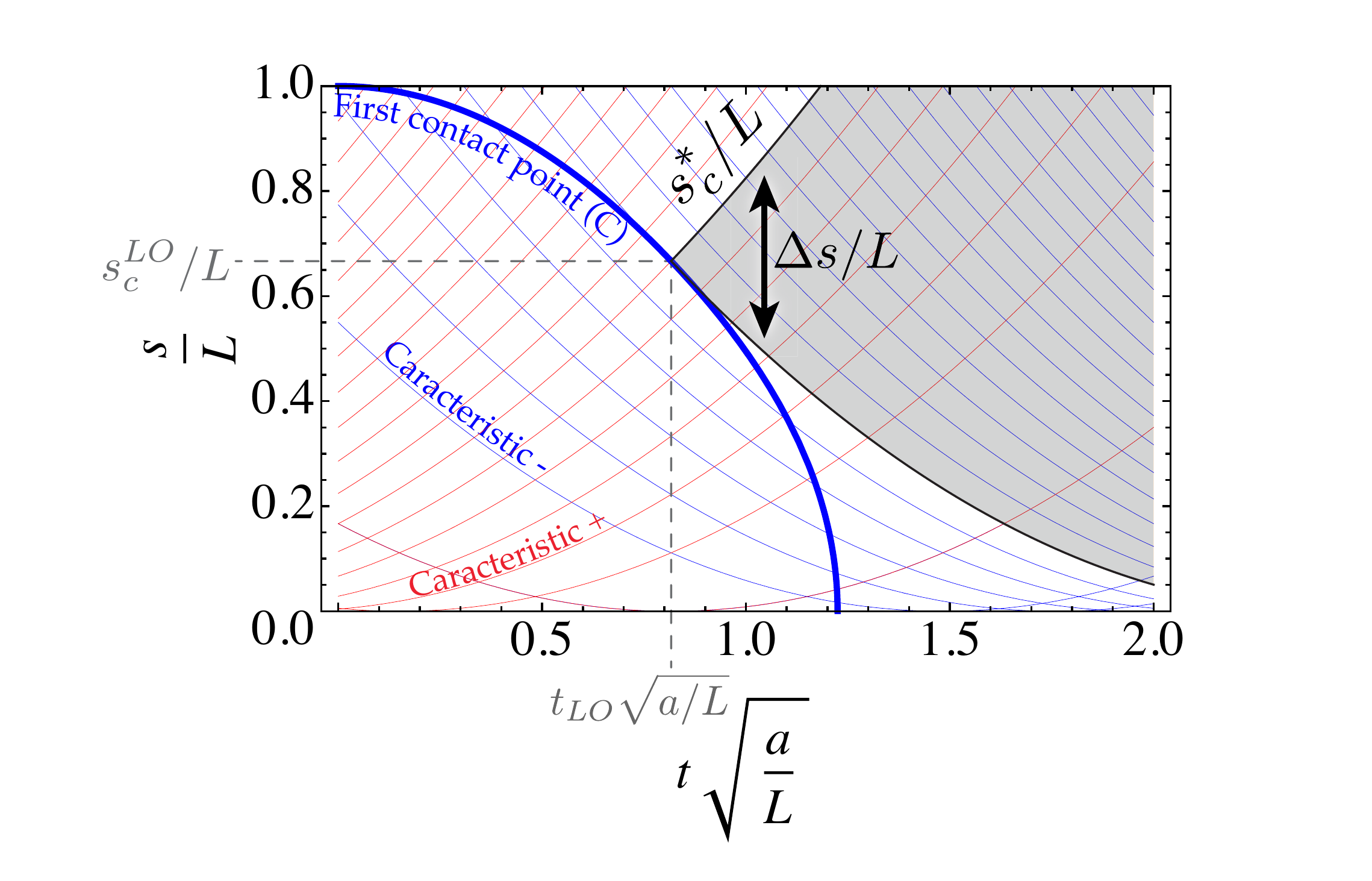}
\caption{ The parameter space $(t\sqrt{a/L},s/L)$ is mapped with the characteristics of the system, color coded as functions of the direction of propagation (given by the sign of $\upd s/\upd t$ in this plot). The time evolution of the position of the contact point $C$, obtained analytically before $\tlift$ and numerically for $t>\tlift$, is shown by solid blue curve. The envelope of characteristics emitted from $C$ at $t=\tlift$ is highlighted by the shaded area. (The simulation timestep is $\Delta t=15\times10^{-4}$
    and the mesh size is $\Delta s = 6.3\times10^{-3}$.  Imposed
    acceleration at endpoint is $a=0.5$, $g=0$, $R=1$ and $L=4\pi$).  }
\label{caractfig}
\end{center}
\end{figure}

We now proceed to explain the size of the region of lift-off as a function of time. We show that the point of contact $C^*(t)$, as defined in fig~\ref{hsm}b, follows a characteristic of the wave equation describing the evolution of the small perturbations to the motion of the string. In other words, the motion of $C^*(t)$ can be viewed as a traveling front produced by the initial lift-off at $\tlift$. To do so, we now consider the length of chain $\Delta s$ that makes up the `ballooning' arch  forming after lift-off. This length is defined to be $\Delta s(t)=s_c^*-s_c$, where $s_c$ and $s_c^*$ correspond to the arc-length coordinate of the contact points $C$ and $C^*$, respectively (see figure \ref{hsm}b).

For times close to $\tlift$, lift-off takes place but it is mild; as a result, this lift-off can be described by $(i)$ extending beyond $\tlift$ the solution maintaining contact with the upper half of the pulley (base solution) and $(ii)$ perturbing it to remove any negative contact pressure. The equations governing the perturbation are obtained by a standard linearization, and are linear wave equations. In these wave equations, the wave speed $c=\sqrt{T/\rho_\ell}$ is both time and space-dependent (recall that $T=T(s,t)$).  Assuming that the tension remains close to its value prior to lift-off{, i.e.~$T\approx\rhol as$} as derived in eq.\eqref{eq:tension}, we  have $c=\sqrt{as}$, and the characteristics of the wave equation satisfy 
\beq
\label{eq:cspeed}
\frac{\upd s}{\upd t}=\pm\sqrt{as}.
\eeq
yielding the expansion waves depicted in figure \ref{caractfig}. 

\begin{figure}
\begin{center}
\includegraphics[width=\textwidth]{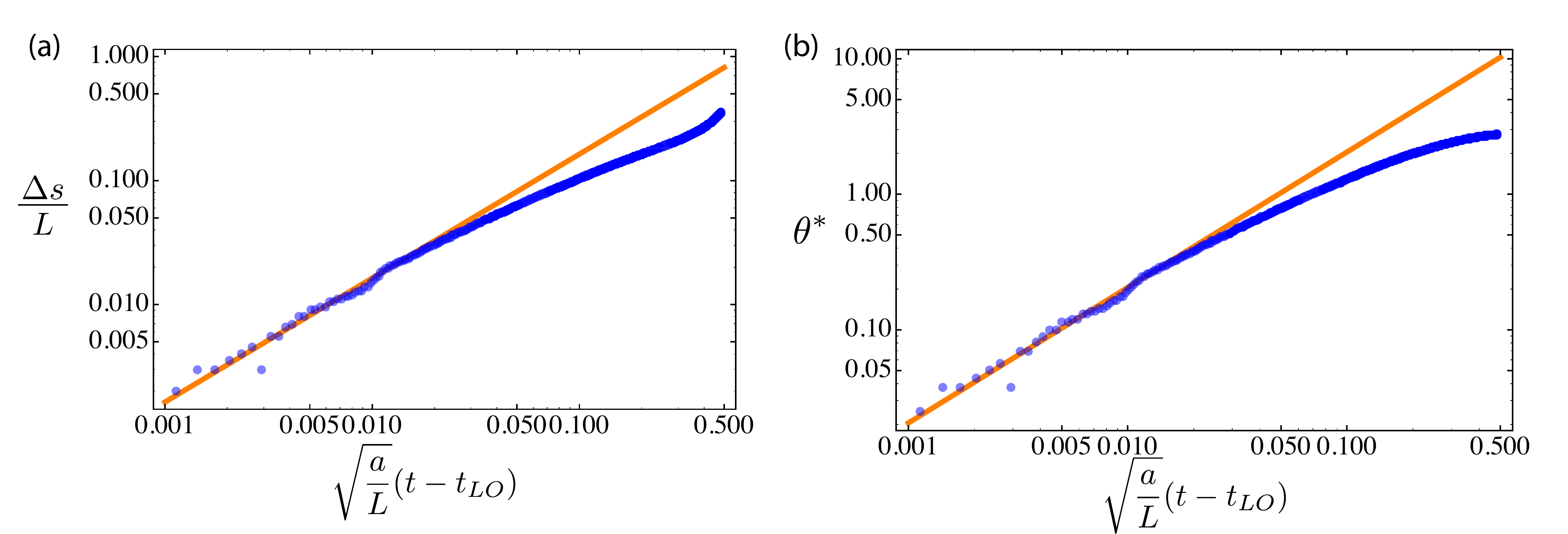}
\caption{Arch structure for $\pi_3=0$. (a) The evolution of $\Delta s/L$ {with time following
lift-off} determined numerically (blue dots) and compared to the
theoretical expression in~\eqref{caract} (orange solid line).  (b)
Test of the prediction $\theta^*\sim \Delta s/R$ derived at the end
of~\S\ref{after}. In both cases the numerical data corresponds to a simulation with timestep $\Delta t=15\times10^{-4}$ and the mesh size $\Delta s = 6.3\times10^{-3}$.  Imposed
    acceleration at endpoint is $a=0.5$, $g=0$, $R=1$ and $L=4\pi$.
}
\label{deltas}
\end{center}
\end{figure}

When lift-off occurs, the chain is locally freed from the pulley at the point $\sclift$ and this information propagates within the envelope formed by the two characteristics that depart from this point in the $(t,s)$ plot. Within this region, the tension in the chain is altered, i.e. the tension of the true solution does not match exactly that predicted by the base solution that ignores lift-off; beyond this region, information can only propagate at the wave speed at the edges, which is, by construction, given by the tension \emph{just} before lift off. The limited propagation speed of information about lift-off  results in the formation of an arch from the point $C$ to a second contact point $C^*$ (see figure \ref{hsm}b).  We hypothesize that the length of the arch $\Delta s(t)=s_c^*-s_c$ is simply the width of the envelope formed by the two characteristics that originate from $C$ at $t=\tlift$ (see figure \ref{caractfig}). Deriving $\Delta s$ from eq.\eqref{eq:cspeed} yields
\beq
\Delta s =2\sqrt\frac{2}{3}\sqrt{aL}(t-\tlift).
\label{caract}
\eeq
We  use numerical simulations to test the estimate~(\ref{caract}). As shown in figure \ref{deltas}a the agreement between the two is asymptotically correct immediately after lift-off (with the correct prefactor). This indicates that the propagation of lift-off front is indeed limited by the rate at which the perturbations can travel through the string. This argument correctly accounts for the short times dynamics, $\Delta s\sim (t-\tlift)$. 

We also note that $ \theta^*$, the angle from $C$ to $C^*$ defined in
figure \ref{caractfig}b, follows the same scaling law and agrees well
with equation \eqref{caract}.  Indeed, recall that $\Delta L =\Delta s
- R \theta^*$; for $t-\tlift\ll1$, we have shown $\Delta L \sim
(t-\tlift)^3$ in equation \eqref{approx} and $\Delta s \sim
(t-\tlift)$ above, it follows that $\Delta s$ and $R \, \theta^*$ must
be equal to leading order in $(t-\tlift)$, \emph{i.e.}~$\theta^*\sim \Delta
s/R$ as checked numerically in figure~\ref{deltas}b.


\section{Snapping}

\begin{figure}
\begin{center}
\includegraphics[width=\textwidth]{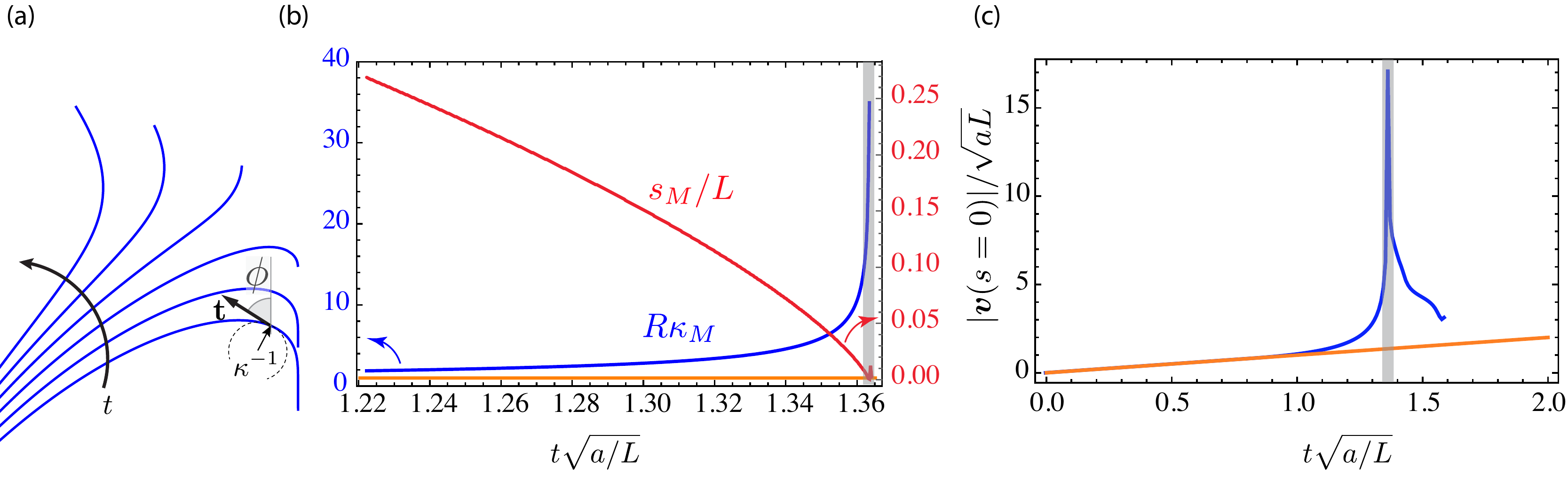}
\caption{Snapping for $\pi_3=0$:  (a) Shapes of the chain right before and after snapping
occurs.  The angle $\phi(s,t)$ separates the tangent $\ve t(s,t)$ and
the the upward pointing vertical axis so that $\kappa(s,t)= \partial
\phi(s,t)/\partial s$, where $\kappa$ is the curvature of the chain.
(b) The maximal curvature of the chain $\kappa_M$ (attained at
position $s_M$) as a function of time and (c) the velocity of the free
end of the chain compared to the {constant acceleration} that drives
the motion.  The grey bands indicate the snapping time.  (The simulation timestep is $\Delta t=15\times10^{-4}$
    and the mesh size is $\Delta s = 6.3\times10^{-3}$.  Imposed
    acceleration at endpoint is $a=0.5$, $g=0$, $R=1$ and $L=4\pi$)}
\label{snap}
\end{center}
\end{figure}

We now describe the kinematics of the free end during snapping.  Let
$\phi(s,t)$ be the angle between the tangent $\ve t$ and the upward
pointing vertical axis.  We define the curvature $\kappa(s,t)=
\partial \phi(s,t)/\partial s$ (recall that the free end corresponds
to $s=0$).  We define snapping as a sudden change in maximal curvature
of the chain as depicted in figure \ref{snap}a-b,
which shows the time evolution of the largest curvature of the chain,
$\kappa_M$.  We observe that the arc-length position $s_M$ at which
this maximum curvature is observed travels towards the free end of the
chain as snapping approaches, as shown in figure~\ref{snap}b.  During
snapping, $\kappa_M$ has a singular behavior as well: the curvature
strongly increases and changes sign, as shown in the last three
configurations in figure \ref{snap}{a} and in figure~\ref{snap}b.
Note that for real ropes this singularity will be regularized by their
finite bending stiffness~\footnote{For chains made up of rigid links,
this singularity is regularized too, by the contact between successive
chain links.}.  Therefore, the singular behavior predicted by our
string model may be seen as the 'outer solution' of a regularized
model.  Note that snapping is associated with a rapid increase of the
free end velocity as illustrated in figure \ref{snap}c.  These
features are similar to what occurs in a cracking
whip~\cite{goriely2002shape} and a falling
chain~\cite{schagerl1997paradox,tomaszewski2005dynamics,tomaszewski2006motion,Grewal2011}.

Let us denote by $t^*$ the time of snapping, defined as the time where
the curvature diverges near the free end.  For optimal accuracy, we
analyze the snapping dynamics based on simulations having a relatively
large value of $\pi_2=R/L=\frac{1}{2\pi}$.  This allows us to use a
fine and uniform spatial discretization of the string.  To ease the
comparison between this specific simulation and the rest of our
results, we introduce a non-dimensionalisation with respect to the
space and time scales of the problem, namely $L$ and $\sqrt{L/a}$.
Dimensionless quantities are denoted by a bar, such as $\overline{t}^*
= t^* / \sqrt{L/a}$.  The dimensionless snapping time $\overline t^*$
is a function of $\pi_2$ and $\pi_3$; in the forthcoming analysis, we 
focus on the particular case $\pi_{2}=1/(2\pi)$ and $\pi_{3} = 0$.

To measure the divergence of curvature from the simulations, we use a
curvature norm that is intended to capture the largest value of
curvature,
\begin{equation}
    \overline K(t) = \left(
    \int_{0}^1
    \left(\overline \kappa(\overline s, \overline t\,)\right)^p\,\mathrm{d}\overline s
    \right)^{\frac{1}{p-1}}
    \textrm{,}
    \label{eq:K}
\end{equation}
where $p\geq 2$ is an integer.  The value of the integer $p$ that is
used results from a trade-off: one should use values of $p$ as large
as possible, and $\overline K$ will then accurately capture the
maximum curvature; on the other hand, if $p$ is too large then
$\overline K$ fluctuates significantly as a result of discretization
errors.  In the following, we take $p=6$; we have verified that the
results are unchanged for $p=8$.

We begin by assuming that as the singularity at $\overline t=\overline
t^*$ is approached, the curvature diverges according to a power law
$\overline \kappa\sim (\overline t^*- \overline t)^{-\alpha}$, for
some exponent $\alpha$ to be determined, on a region of size $\sim 1/
\overline \kappa$.  The region of divergence contributes an amount
$\sim( \overline t^*- \overline t)^{-(p-1)\,\alpha}$ to the integral
in equation~(\ref{eq:K}): as soon as $\alpha(p-1)>0$, this
contribution makes the integral diverge at the time of snapping
$\overline{t}^*$ and $\overline{t}^*$ can be identified from a
numerical plot of $\overline{t}$ as a function of time---the condition
$\alpha(p-1)>0$ is indeed satisfied as $p=6$ and $\alpha>0$, see
below.  One can check that the quantity $\overline K(t)$ diverges as
$\overline K\sim (\overline t^*-\overline t)^{-\alpha}$: a numerical
plot of $\overline{K}$ versus $\overline{t}$ yields the exponent
$\alpha$ as well, independently of the particular value of $p$ chosen
in the definition of $\overline{K}$.

\begin{figure}
    \centerline{\includegraphics[width=.95\textwidth]{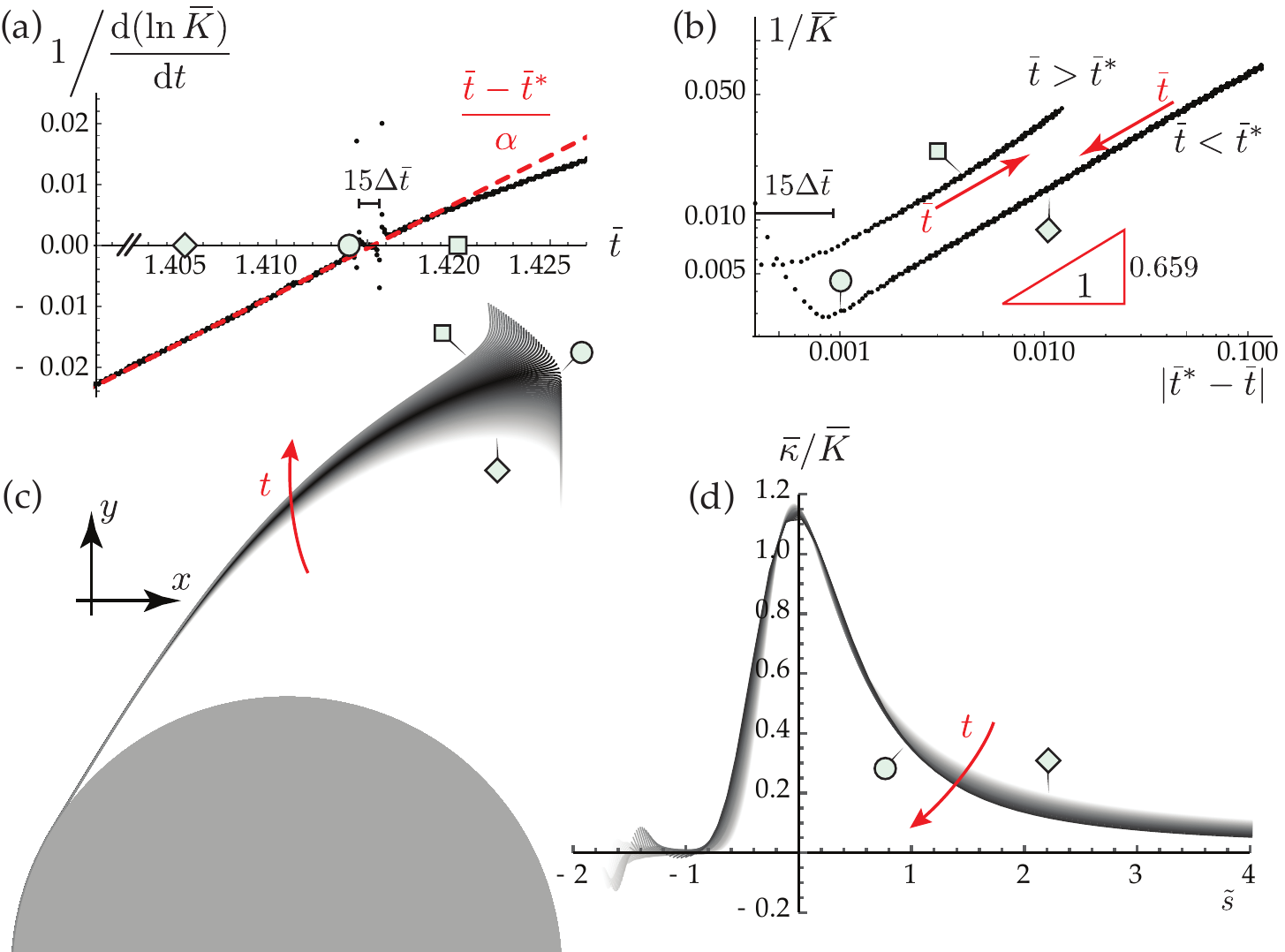}}
    \caption{ Searching for self-similar solutions in the snapping (for this
    particular simulation, the linear mass is $\rhol = 5$ and the
    radius of pulley is $R=1$ and $L=2\pi$ such that
    $\pi_2=R/L=1/(2\pi)$.  There is no gravity and no bending
    rigidity.  The simulation timestep is $\Delta t=2\times10^{-4}$
    and the mesh size is $\Delta s = 3.14\times10^{-3}$.  Imposed
    acceleration at endpoint is $a=0.5$). (a) Identifying the singularity exponent $\alpha$ for the
    curvature and the critical time $\overline t^*$ using the quantity
    $\overline K(\overline t)$ defined in \eqref{eq:K}.  From our
    numerical simulations we find that
    $(\alpha,\overline{t}^*)=(0.659,1.41531)$ (red dashed line).  (b)
    These values may be checked in a standard log-log plot.  The wells
    are numerical artifacts: they correspond to values of $|\overline
    t^*-\overline t|$ as small as a few simulation steps, implying
    that the curvature is very peaked and prone to discretization
    errors.  (c) Configurations right before and after the snapping
    time, $1.405\leq \overline t \leq 1.42$ ($\overline
    t^*\approx 1.415$): darker configurations correspond to times
    closer to snapping and reference configurations are denoted by the
    symbols (\protect\tikz \protect\draw[black,fill=PineGreen!10]
    (-.11,0) -- (.0,.11) -- (.11,0)-- (0,-.11)--(-.11,0);),
    (\protect\tikz \protect\draw[black,fill=PineGreen!10] (0,0) circle
    (.75ex);) and (\protect\tikz
    \protect\draw[black,fill=PineGreen!10] (-.09,-.09) -- (.09,-.09)
    -- (.09,.09)-- (-.09,.09)--(-.09,-.09);) (d)~Collapse of the
    curvature in rescaled variables $(\tilde s, \overline
    \kappa/\overline K)$ using the same shading convention for $1.405\leq \overline
t \leq1.414$ (before snapping occurs).}
    \protect\label{fig:mmtkaFits}
\end{figure}

The snapping time $\overline{t}^*$ and the exponent $\alpha$ entering
in the power law $\overline{K}(t)\sim |t-t^*|^{-\alpha}$ are best
determined from the simulations by plotting the quantity
\begin{equation}
\left(\frac{\mathrm{d}(\ln \overline K)}{\mathrm{d} \overline t}\right)^{-1}\sim \frac{\overline t^*-\overline t}{\alpha}
\end{equation}
as a function of $\overline t$, see figure~\ref{fig:mmtkaFits}a.

The points collapse onto a straight line, aside from a thin band of
width $\approx 15 \Delta \overline t$ where the large curvature is not
well resolved by the discretization.  A linear fit in
figure~\ref{fig:mmtkaFits}a yields both the snapping time $\overline
t^*$ (where the fitting line crosses the $t$-axis) and the exponent
$\alpha$ (the reciprocal of the slope).  This fit yields a value of
the exponent $\alpha = 0.659$ which is numerically close to 2/3 (we
do not know whether this is a coincidence) and $\overline t^*=
1.415$.  The latter is close to (but less than) the singular time
$2\tlift \sqrt{a/L}=2\sqrt{2/3}\approx 1.63 $ found by extrapolating
the asymptotic behavior in the immediate aftermath after lift-off, see
equation~(\ref{eqn:vhanggeq0}).  These numerical values of
$\overline{t}^*$ and $\alpha$ have been obtained in the particular
case $\pi_{2}=1/(2\pi)$ and $\pi_{3} = 0$.  We suspect, however, that
the exponent $\alpha$ can be explained by a boundary-layer theory and
that it is actually independent on $\pi_{2}$ and $\pi_{3}$.
Unfortunately, we can offer no proof for this statement.

As observed in the experiments (see \S\ref{phenomenon}), the string
remains straight in the simulations along an interval of length
$\ell(t)$ comprising the free end, at all times until $t^*$.  This
straight portion of the string is apparent in figures~\ref{snap}a
and~\ref{fig:mmtkaFits}c.  It defines the region denoted by (I), which
is adjacent to the region (II) of large curvature (see
Fig~\ref{fig:DVR}c).  The dimensionless length $\overline{\ell} =
\ell/L$ is found numerically to scale like $\overline \ell\sim
(\overline t^{*}- \overline t)^\beta$, with $\beta = 1.08\pm0.2$ (fit
not shown).  The numerical accuracy on this exponent $\beta$ is not as
good as that of $\alpha$ as the determination of the straight region
is sensitive to numerical noise.  It is clear from our numerical data,
however, that the exponent $\beta$ is larger than $\alpha$, meaning
that the length of the straight region near the free end vanishes more
quickly than the {minimum} radius of curvature as $\overline t\to
\overline t^*$, $\overline t<\overline t^*$.

Next, we proceed to use our analysis to uncover a self-similar
behavior in the curvature profile.  Anticipating that the point of
maximal curvature plays a central role we introduce a weighing to
capture its position:
\begin{equation}
    \langle \overline s\rangle({\overline t}) = \frac{
    \int_{0}^1
   \overline s\,
  \overline  \kappa^p(\overline s,\overline t)\,\mathrm{d}\overline s
    }{
    \int_{0}^1
  \overline  \kappa^p(\overline s,\overline t)\,\mathrm{d}\overline s
    }
    \textrm{.}
    \label{eq:weightedAverage}
\end{equation}
By design, the weighting is concentrated in the region of large
curvature: $\langle \overline s\rangle({\overline t})$ yields the
typical value of $\overline s$ at time $\overline t$ in the region
(II) where the curvature diverges.

Finally we rescale the arc-length parameter to
\begin{equation}
    \tilde {s} = \frac{\overline s - \langle \overline s\rangle({\overline t})}{1/\overline K(\overline t)}
    \textrm{,}
    \label{eq:rescaledArcLength}
    \end{equation} 
By design, $\tilde{s}=0$ lies in the centre of the region with high
curvature.  In this definition, the offset $\langle \overline
s\rangle({\overline t})$ allows us to ignore the straight region
entirely (which shrinks according to a different exponent).  We
consider a set of simulation snapshots such that $1.405\leq \overline
t \leq1.414$, {i.e.}~such that $0.001\leq \overline t^*-\overline
t\leq 0.01$.  This corresponds to the configurations bounded by the
symbols (\protect\tikz \protect\draw[black,fill=PineGreen!10] (-.11,0)
-- (.0,.11) -- (.11,0)-- (0,-.11)--(-.11,0);) and (\protect\tikz
\protect\draw[black,fill=PineGreen!10] (0,0) circle (.75ex);) in
figure \ref{fig:mmtkaFits}b-c, with darker plots corresponding to
times closer to the snapping time $\overline t^*$. 

In figure \ref{fig:mmtkaFits}d, we plot the rescaled curvature
$\overline\kappa(s,t)/\overline K(t)$ as a function of the rescaled
arc-length $\tilde{s}$ and obtain a good collapse.  Note that the
unscaled maximum curvature $\overline K$ varies by a factor $\sim 5$,
from approximately 65 to 308, between the first and last snapshot in
this series.  Therefore, the collapse shows that the curvature
distribution is self-similar close to the snapping time $\overline
t^*$: consistently with our initial scaling assumption, the curvature
scales like $\overline \kappa\sim (\overline t^*-\overline
t)^{-\alpha}$ in a region of size $\overline s - \langle
\overline{s}\rangle\sim(\overline t^{*}-\overline t)^\alpha$.  Note that this analysis focuses on the behavior of the chain prior to snapping, so that $\overline t < \overline t^*$. 

In summary, our analysis shows that the snapping singularity may
characterized by a self-similar solution, whose scaling behaviours
have been identified numerically.

\section{Conclusions} 
In this paper, we have considered a degenerate version of Atwood's machine, in which a single mass pulls a chain around a pulley.  In stark contrast with the apparent simplicity of the setup we have found that the dynamics is extremely rich, successively displaying a ballooning instability of the chain and a snapping motion of its free end, reminiscent of what is seen in a cracking whip. We have shown that  the chain dynamics is well captured by a frictionless string model and that some of its features may be captured by simple arguments.  In particular we have shown that the geometry of the problem, through the imposed rotation of the chain around the pulley, is key to understanding how the end of the chain is able to 'beat' the free-fall that drives its motion.

Our observations can be used to speculate on the peculiar hunting techniques of a variety of amphibians. Indeed, instead of throwing their tongue in a straight motion (as observed in chameleons~\cite{Anderson:2010wm,chameleon}), certain species of toads~\cite{Deban:2011ve} and salamanders~\cite{Deban:2007td}  adopt an unfurling tongue strategy. Of course, the reasons for such a mechanism are many and varied but we believe that the increase of tip velocity observed in the case of a chain is  likely to reappear in problems involving a finite bending stiffness. It is then natural to wonder whether this geometrical amplification of acceleration may be  used by these amphibians to allow them to maximize their chances of capturing a prey?

Finally, we note that while we have been able to rationalize some of
the observations from experiment and simulation, others remain
elusive.  For example, predicting the shape of the ballooning region
either with simple arguments, or preferably analytically, remains
beyond our reach.  Difficulties in doing so arise from the fact that
the `base solution' of the problem is unsteady (since $a>0$).  An
interesting avenue of research would be to explore the case where the
chain is pulled at constant speed thus without any acceleration,
potentially allowing for analytical developments.  Similarly, there is
hope that the self-similar exponent $\alpha$ and $\beta$ identified in
the previous section can be explained by some boundary layer theory in
future work.

\section*{Acknowledgment}

This publication is based in part upon work  partly funded by the European Research Council (ERC) under the European Union's Horizon 2020 research and innovation programme (grant agreement No 637334, GADGET to DV).

\appendix
{
\section{The effect of gravity immediately after lift-off \label{sec:grav}}

For simplicity, the main portion of the paper focusses on the problem in the absence of gravity. In this Appendix, we show how the results derived in \S\ref{after}\ref{sec:aftervel} for the behaviour immediately following lift-off are altered once gravity is included. These results confirm that, in this case at least, gravity affects the results quantitatively, rather than qualitatively.

\subsection{The unknown chain acceleration}

Immediately after lift-off, the free end of the chain accelerates at an unknown rate $\ahang$. To determine this acceleration, we repeat the argument of \S~\ref{after}\ref{sec:aftervel} incorporating gravity: the normal force arising from the tension at the material position that was at C when lift-off started ($\rhol (\ahang+g)\sclift\kappa$)  is equated with the force arising from the centripetal acceleration ($\rhol \vhang^2\kappa$) for some undetermined curvature $\kappa$. We then have that \eqref{eqn:HSAnew} becomes
\beq
(\ahang+g)\sclift=\vhang^2,
\label{eqn:HSAnew:gneq0}
\eeq which can be solved with initial condition $\vhang(\tlift)=a\tlift$ to give
\beq
\vhang=\sqrt{g\sclift}\coth \left[\alpha-\sqrt{\frac{g}{\sclift}}(t-\tlift)\right],
\label{eqn:vhangcoth}
\eeq where
\beq
\coth\alpha=\frac{a\tlift}{\sqrt{g\sclift}}.
\eeq The form of \eqref{eqn:vhangcoth} appears to be substantially different to that in the absence of gravity, \eqref{eqn:vhanggeq0}. However, the qualitative behaviour is, in fact very similar: $\vhang(t)$ is an increasing function of $t>\tlift$ and has a singularity at
\beq
t=\tlift+\sqrt{\frac{\sclift}{g}}\alpha.
\eeq Furthermore, a Taylor expansion of \eqref{eqn:vhangcoth} reveals that
\beq
\vhang=at+\frac{a^2\tlift}{\sclift}(t-\tlift)^2+O(t-\tlift)^3,
\eeq so that the change in the pulling velocity occurs at $O(t-\tlift)^2$, as in the case without gravity (see figure \ref{hsm}a).

\subsection{The excess length $\Delta L$}

From the expression \eqref{eqn:vhangcoth} for the acceleration of the hanging portion of the chain, we find that the excess length absorbed by the lifted off portion is
\begin{align}
\Delta L&=-\sclift\log\frac{\sinh \left[\alpha-\sqrt{\frac{g}{\sclift}}(t-\tlift)\right]}{\sinh\alpha}+\frac{a}{2}(\tlift^2-t^2)\nonumber\\
&= \left(a+\tfrac{g}{3}\right)\frac{(t-\tlift)^3}{3\tlift}+O[(t-\tlift)^4].
\end{align} This cubic growth of the excess length with time following lift-off echoes that found in the absence of gravity, given by \eqref{approx}. As already seen, the only significant difference is in the prefactor.

}

\end{document}